\documentclass[preprint, showpacs, aps, draft]{revtex4}
\usepackage{bm}
\begin{document}
\title{Exact uncertainty properties of bosonic fields}
\author{Michael J. W. Hall}
\author{Kailash Kumar}
\affiliation{Theoretical Physics, IAS, \\ Australian National
University,\\
Canberra ACT 0200, Australia}
\author{Marcel Reginatto}
\affiliation{Physikalisch-Technische Bundesanstalt, \\
Bundesallee 100\\
38116 Braunschweig, Germany}
\begin{abstract}
The momentum density conjugate to a bosonic quantum field splits naturally
into the sum of a classical component and a nonclassical component.  It
is shown that the field and the nonclassical component of the momentum
density satisfy an {\it exact}
uncertainty relation, i.e., an equality, which underlies 
the Heisenberg-type uncertainty relation for fields.
This motivates a new approach to deriving and interpreting bosonic 
quantum fields, based on an exact uncertainty principle.  In particular,
the postulate that an ensemble of classical fields is subject to
nonclassical 
momentum fluctuations, of a strength determined by the field
uncertainty, leads from the classical to the quantum field equations.
Examples include scalar, electromagnetic and gravitational fields.  For
the latter case the exact uncertainty principle specifies a unique
(non-Laplacian) operator ordering for the Wheeler-deWitt equation. 
\end{abstract}
\maketitle

\section{Introduction}

The Heisenberg uncertainty relation
\begin{equation} \label{hur}
\Delta x \Delta p \geq \hbar/2
\end{equation}
for the rms position and momentum uncertainties of a quantum particle is
a well known feature of quantum mechanics. It has recently 
been shown, however, that this relation is a consequence of a more
fundamental connection between the statistics of complementary quantum
observables.

In particular, the distinguishing ``nonclassical'' 
property of complementary observables is
that they cannot be simultaneously measured to an arbitrary accuracy.
It is therefore natural to consider the decomposition of one such
observable, the momentum say, into the sum of a ``classical'' and a
``nonclassical'' component: 
\[ \hat{p} = \hat{p}^{cl} + \hat{p}^{nc}~, \]
where the classical component, $\hat{p}^{cl}$, is defined as that observable
{\it closest} to $\hat{p}$ (in a statistical sense) which {\it is} 
simultaneously
measurable with the complementary observable $\hat{x}$ \cite{eur}.  

It turns out that
such decompositions do indeed, in a number of ways, 
neatly separate classical and
nonclassical properties of quantum observables.  
For example, for one-dimensional particles the {\it
nonclassical} component of the momentum satisfies the uncertainty
relation 
\begin{equation} \label{eur}
\delta x \Delta p^{nc} = \hbar/2
\end{equation}
for {\it all} pure states, where $\delta x$ denotes 
a measure of position uncertainty from
classical statistics called the Fisher length \cite{eur, hallfish}.
This {\it exact} uncertainty relation is far stronger than
(and implies) the corresponding Heisenberg uncertainty relation in Eq.
(\ref{hur}). 

The surprising fact that quantum systems satisfy exact uncertainty
relations has recently provided the basis for {\it deriving} much of the quantum
formalism, from an exact uncertainty principle.  In particular, the
assumption that a classical ensemble is subjected to nonclassical momentum
fluctuations, of a strength inversely proportional to uncertainty in
position, has been shown to lead directly from the classical equations
of motion to the Schr\"{o}dinger equation \cite{hallreg}. A brief
overview of exact uncertainty properties of quantum particles is
provided by 
the conference paper in Ref.~\cite{bamberg}. 

The aim of this paper is to show that the abovementioned results
can be generalised to bosonic quantum fields.  
The paper is therefore divided into two separate and logically distinct
sections:  exact uncertainty relations in Sec. II, and an exact
uncertainty principle in Sec. III.  Each of these sections may be read
without reference to the other, according to the reader's interest.

In Sec. II it is shown 
that the statistics of {\it any} bosonic field $\hat{f}$ and its conjugate
momentum field $\hat{g}$ are connected by an exact uncertainty relation
analogous to Eq. (\ref{eur}). This uncertainty relation provides a
precise equality between the statistical 
covariance of the non-classical component 
of $\hat{g}$ and the Fisher covariance of $\hat{f}$, and underlies a
Heisenberg-type inequality relating the statistical covariances of $\hat{f}$ and
$\hat{g}$.  

In Sec. III  
it is shown that the
quantum field equations follow from an exact uncertainty principle, for
the case of bosonic fields with Hamiltonians {\it quadratic} in the field
momenta (eg, scalar, electromagnetic and gravitational fields).  
This ``exact uncertainty'' approach is extremely minimalist in nature:
unlike canonical quantisation, it does
not use nor make any assumptions about
the existence of operators, Hilbert spaces, complex
amplitudes, inner products, 
linearity, superposition, or the like. The sole ``nonclassical''
element needed is the addition of fluctuations to the
momentum density of members of a classical ensemble of fields, 
with the fluctuation statistics assumed to be determined by the ensemble
statistics. The exact uncertainty approach is thus 
conceptually very simple, being based
on the core notion of statistical uncertainty 
(intrinsic to any interpretation of quantum theory). 

As a bonus, the
exact uncertainty approach further
implies a unique operator ordering for the quantum field equations -
something
which the canonical quantisation procedure is unable to do.  This is
relevant in particular to 
the Wheeler-deWitt equation (where 
the associated unique ordering is, moreover,  
consistent with
Vilenkin's ``tunneling" boundary condition for inflationary cosmology
\cite {vilenkin}).  

It is remarkable
that the basic underlying concept - the addition of ``nonclassical'' momentum
fluctuations to a classical ensemble - carries through from quantum
particles to quantum fields, without creating
conceptual difficulties. This is a strength of the exact uncertainty 
approach, not mirrored in other approaches 
that rely on connecting the equations of motion of
classical and quantum ensembles.  For example, the so-called ``causal''
interpretation of Bohm and co-workers 
is explicity non-local \cite{holland} (and hence
{\it non}-causal for relativistic fields), while one cannot
simultaneously define both the electric and magnetic fields in
generalisations of Nelson's stochastic approach to electromagnetic
fields \cite{nelson}. 
  
Necessary elements from
classical field theory and functional analysis are briefly summarised in
the Appendices.

\section{Exact uncertainty relations for fields}

A Heisenberg-type uncertainty relation  may be given 
for scalar fields, in the form of
an inequality for the covariance functions of the
field and its conjugate momentum density.  However, a stronger result
follows via a natural
decomposition of the momentum density into classical and nonclassical
components, and the use of the Fisher covariance function from classical
statistical estimation theory.  In particular,
an {\it exact} uncertainty relation for scalar fields is obtained in
Sec. II.D below, which underlies and implies the corresponding Heisenberg-type
uncertainty relation, and which may be generalised to arbitrary bosonic
fields. 

This derivation of exact uncertainty relations,
i.e., precise quantitative connections between the statistics of
conjugate fields, from the quantum formalism, is to some extent
conceptually reversed in Sec. III, where the quantisation of a large
class of classical fields is achieved via the use of an ``exact
uncertainty'' principle. 
The reader primarily interested in this new 
approach to obtaining the quantum field equations may wish to skip
directly to Sec. III.

\subsection{Heisenberg-type relations}

We begin with the simplest case: a real scalar field $f$ with conjugate
momentum density $g$. Spatial coordinates will be denoted by $x$
(irrespective of dimension), and the values of $f$ and $g$ at position
$x$ by $f_x$ and $g_x$ respectively.  

The underlying origin of the field is not
at issue here - it could be a relativistic field satisfying the
Klein-Gordon equation \cite{brown}, or a nonrelativistic field
describing vibrations of a stretched string \cite{goldstein}.  What
is important for the purposes of this Section is that the corresponding
quantised fields $\hat{f}$ and $\hat{g}$ satisfy the equal-time
commutation relations \cite{brown}
\begin{equation} \label{comm}
[\hat{f}_x , \hat{f}_{x'}] = [\hat{g}_x , \hat{g}_{x'}] = 0,~~~~
[\hat{f}_x ,\hat{g}_{x'}] = i\hbar\delta(x-x') . 
\end{equation}
Thus, in the Schr\"{o}dinger representation, where the
quantum state of the field at a given time
is formally described by a complex amplitude
functional $\Psi[f]$, the action of $\hat{f}$ and $\hat{g}$ is given by
\cite{schweber}
\begin{equation} \label{srep}
\hat{f} \Psi = f \Psi,~~~~\hat{g} \Psi =
\frac{\hbar}{i}\frac{\delta\Psi}{\delta f} ,
\end{equation}
where $\delta /\delta f$ denotes the usual functional derivative (see
also Appendix A).

The non-vanishing commutator in Eq. (\ref{comm}) reflects the 
complementary nature of $\hat{f}$ and $\hat{g}$:  in general one must
choose to accurately measure {\it either} the field {\it or} its
momentum density. This is analogous to the complementary nature of the
position and momentum observables ${\bf X}$ and ${\bf P}$ of a system of
quantum particles, with non-vanishing commutator
$[X_j,P_k]=i\hbar\delta_{jk}$. In the latter case one has the
corresponding Heisenberg uncertainty relation \cite{cover}
\begin{equation} \label{discrete}
{\rm Cov}({\bf X})~ {\rm Cov}({\bf P}) \geq (\hbar^2/4) I 
\end{equation}
for the covariance matrices of ${\bf X}$ and ${\bf P}$, where $[{\rm
Cov}({\bf A})]_{jk}:=\langle A_jA_k\rangle - \langle A_j\rangle\langle
A_k\rangle$, and $I$ denotes the identity matrix.  Thus, since the
covariance matrix vanishes for the dispersion-free case, the
complementary observables ${\bf X}$ and
${\bf P}$ cannot be simultaneously specified.

The covariance {\it function} of a given
field $h$ is defined by 
\begin{equation} \label{hcov}
\left[{\rm Cov}(h)\right]_{xx'} := 
\langle h_xh_{x'}\rangle -
\langle h_x\rangle \langle h_{x'}\rangle ,
\end{equation}
where $\langle ~\rangle$ denotes an ensemble average.
 Thus the diagonal elements of the covariance
function (i.e., $x=x'$) represent the
variance of the field at each point, and the
off-diagonal elements correspond to the degree of linear correlation
between field values at different points.  

The covariance functions of the conjugate fields $f$ and $g$ satisfy the 
Heisenberg-type inequality 
\begin{equation} \label{hfield}
{\rm Cov}(f) \ast {\rm Cov}(g) \succeq (\hbar^2/4)\openone .
\end{equation}
analogous to Eq.~(\ref{discrete}), 
where we use a matrix-type notation with multiplication $A\ast B$
defined
by
\[
\left[ A\ast B\right]_{x x'} := \int dx'' A_{x x''}B_{x'' x'} ,
\]
multiplicative identity $\openone$ defined by
\[
\openone_{xx'} := \delta(x-x') , \]
and the ordering $A\succeq B$ holds if and only if
\[
h^\dagger (A-B) h := \int dxdx'\,(A-B)_{xx'}h^*_{x}h_{x'} \geq 0 \]
for all $h$. 
Eq.~(\ref{hfield}) is clearly a continuous analogue of the Heisenberg
inequality in (\ref{discrete}), and indeed 
may be ``proved'' from the latter inequality by
a discretization argument (where the spatial degree of freedom is
represented by discrete cells of volume $\epsilon$ at positions $x_j$, 
the identifications $X_j=f_{x_j}$, $P_j=\epsilon g_{x_j}$,
$\epsilon\sum_j=\int dx$ are made, and
the limit $\epsilon\rightarrow 0$ is taken).  A more rigorous proof is
given in Sec. II.D below. 

Eq. (\ref{hfield}) represents a field-theoretic
generalisation of the 1-dimensional Heisenberg uncertainty relation in
Eq. (\ref{hur}).  Moreover, 
it has the same physical significance: since
the covariance function of field $h$ vanishes for the
dispersion-free case, the conjugate fields 
$f$ and $g$ cannot simultaneously be specified. It will be shown in Sec.
II. D that this inequality can be replaced by an {\it exact }uncertainty
relation analogous to Eq. (\ref{eur}).  However, two concepts must first
be introduced:  decomposition of the quantum momentum density into ``classical''
and ``nonclassical'' components, and the Fisher covariance function. 

\subsection{Decomposition of the momentum density}
Since, in contrast to classical fields, the quantum momentum density
cannot be measured simultaneously with the field, we define the {\it
classical component} of the momentum density to be that observable which
is {\it closest} to $\hat{g}$, statistically speaking, 
under the constraint of being
comeasurable with $\hat{f}$.  Thus measurement of the classical component
provides the {\it best possible} estimate of the momentum density 
compatible with a measurement of the field.

More formally, let $\hat{g}^{cl}$ denote the classical component of the
momentum density $\hat{g}$.  The comeasurability of $\hat{f}$ and
$\hat{g}^{cl}$ then implies that the action of the latter in the
representation of Eq. (\ref{srep}) has the form
\begin{equation} \label{cond1}
\hat{g}^{cl} \Psi = g^{cl}[f]\Psi
\end{equation}
for some real functional $g^{cl}[f]$.   
Further, for the classical component to be as close as possible
to the quantum momentum density, the average
deviation of $\hat{g}^{cl}$ and $\hat{g}$ must be as small as possible,
i.e., 
\begin{equation} \label{cond2}
\langle ~(\hat{g}_x - \hat{g}^{cl}_x)^2~ \rangle_\Psi = {\rm
minimum}
\end{equation}
at each position $x$.

Conditions (\ref{cond1}) and (\ref{cond2}) determine the classical
component of the momentum density uniquely, for each quantum state
$\Psi$: one finds, as shown further below, that  
\begin{equation} \label{gcl}
g^{cl}[f] = \frac{\hbar}{2i} \left[\frac{1}{\Psi}\frac{\delta\Psi}{\delta f} 
- \frac{1}{\Psi^*}\frac{\delta\Psi^*}{\delta f} \right]. 
\end{equation}

Note that the classical component depends on
$\Psi$ - to make the best possible estimate one needs 
some knowledge about the state of the field \cite{footest}.  
Thus one should, strictly speaking, use a notation such as
$\hat{g}^{cl,\Psi}$ for the classical component, particularly
if one wishes to calculate expectation
values such as $\langle\hat{g}^{cl,\Psi}\rangle_\Phi$ for states $\Phi$
other than $\Psi$.  However, since quantities involving the classical
component will in fact only
be evaluated for the corresponding state $\Psi$ in what follows, such 
explicit notational dependence on the state may be conveniently dispensed with. 

Note also (for each state $\Psi$) that, since the operator $\hat{g}^{cl}$ 
commutes with the field operator
$\hat{f}$, the former operator is measurable to the
same degree as the latter. 
In particular, the experimentalist's in-principle procedure for measuring
$\hat{g}^{cl}$ on state $\Psi$ 
is to (i) prepare the quantum state $\Psi$; (ii) measure
the field $\hat{f}$; and (iii) for measurement result $f$ for $\hat{f}$
calculate the corresponding measurement result $g^{cl}[f]$ for $\hat{g}^{cl}$. 

It follows that the momentum density of a quantum field admits a natural
(state-dependent) decomposition,
\begin{equation}
\label{gdecomp} \hat{g} = \hat{g}^{cl} + \hat{g}^{nc} ,
\end{equation}
into the sum of a classical component $\hat{g}^{cl}$,
corresponding to
that part of $\hat{g}$ which is comeasurable with $\hat{f}$, and a {\it
nonclassical} component $\hat{g}^{nc}$, corresponding to an 
intrinsically ``quantum'' remainder. Further, these
components satisfy the relations
\begin{equation} \label{gav}
\langle \hat{g}_x\rangle = \langle\hat{g}^{cl}_x\rangle,
~~~~\langle\hat{g}^{nc}_x\rangle = 0,
\end{equation} 
\begin{equation} \label{gvar}
(\Delta g_x)^2 = (\Delta g^{cl}_x)^2 + (\Delta g^{nc}_x)^2 , 
\end{equation}
and hence the classical and nonclassical components also represent an
``average'' part and a ``fluctuation'' part respectively, the 
uncertainties of which add in quadrature. 

To derive the explicit form in Eq. (\ref{gcl}) for the classical
component, consider any other field operator $\hat{h}$ comeasurable
with $\hat{f}$, so that $\hat{h}\Psi = h[f]\Psi$ in analogy to Eq.
(\ref{cond1}).  Then from Eqs. (\ref{srep}) and (\ref{gcl}), and Eq.
(\ref{div}) of Appendix A (assuming $\langle h\rangle$ is finite), one has
\begin{eqnarray*}
\langle \hat{g}_x\hat{h}_x + \hat{h}_x\hat{g}_x\rangle & =
&\frac{\hbar}{i}\int D\! f\, \Psi^*\left[\frac{\delta (h_x\Psi)}{\delta
f_x} + h_x\frac{\delta\Psi}{\delta f_x}\right]\\
& = & (\hbar/i)\int D\! f\, [\Psi^*(\delta\Psi/\delta f_x) - \Psi
(\delta\Psi^* /\delta f_x)]h_x\\
& = & 2\langle \hat{g}^{cl}_x \hat{h}_x\rangle ,
\end{eqnarray*}
where $\langle\hat{A}\rangle$ is defined as the functional integral
$\int D\! f\, \Psi^*\hat{A}\Psi$ (see Appendix A).  Hence
\begin{eqnarray*}
\langle (\hat{g}_x-\hat{h}_x)^2\rangle & = & \langle
(\hat{g}_x)^2\rangle + \langle(\hat{h}_x)^2\rangle
-2\langle\hat{g}^{cl}_x\hat{h}_x\rangle\\
& = & \langle(\hat{g}_x)^2\rangle - \langle(\hat{g}^{cl}_x)^2
\rangle + \langle (\hat{h}_x - \hat{g}^{cl}_x)^2\rangle .
\end{eqnarray*}
Since the last term on the right is nonnegative, the lefthand side is
minimised for the choice $\hat{h}\equiv\hat{g}^{cl}$, as required by
condition (\ref{cond2}).  

Note that Eq.~(\ref{gav}) follows immediately
from Eqs.~(\ref{gcl}) and (\ref{gdecomp}), and Eq.~(\ref{gvar}) then
follows by substituting $\hat{h}\equiv\hat{g}^{cl}$ into the expression 
immediately above.  Eq. (\ref{gvar}) in fact corresponds to the diagonal
elements of the more general covariance relation
\begin{equation} \label{covadd}
{\rm Cov}(g) = {\rm Cov}(g^{cl}) + {\rm Cov}(g^{nc}) ,
\end{equation}
which may be derived by explicit calculation of ${\rm Cov}(g-g^{cl})$.  The
classical and nonclassical field components are therefore linearly uncorrelated. 

\subsection{Fisher covariance}
Many different classical measures of statistical
uncertainty, such as variance and entropy, have been used in writing
Heisenberg-type inequalities for quantum particles \cite{uffink}. 
Similarly, there is a wide degree of freedom in choosing measures of
uncertainty for quantum fields - the covariance function ${\rm
Cov}(f)$ appearing in Eq.~(\ref{hfield}) only represents one particular choice.
An alternative choice is provided by classical statistical
estimation theory:  the {\it Fisher} covariance function.

The Fisher covariance function ${\rm Cov_F}(f)$ is, like the standard
covariance function ${\rm Cov}(f)$, a positive and symmetric two-point 
function which vanishes for dispersion-free ensembles, and is hence a
suitable candidate for describing field uncertainty over an ensemble.
Its main role in classical statistics is as the {\it ``minimum''} covariance
function associated with estimates of translations of the field.

To motivate ${\rm Cov_F}(f)$, consider a classical ensemble of fields
described by some probability functional $P[f]$, with ensemble average $\langle
f\rangle = \int D\! f\,Pf = 0$.  If each member of the
ensemble is translated by some amount $\overline{f}$, a new ensemble is
obtained, described by $P[f-\overline{f}]$ and having field average
$\langle f\rangle = \overline{f}$.  The two ensembles thus
differ only by the displacement field $\overline{f}$, and hence
distinguishing between two such ensembles is equivalent to estimating
this relative displacement.

How easily {\it can} such ensembles be distinguished? 
The basic answer is that 
any {\it unbiased} estimate $\tilde{f}$ of the displacement 
field (i.e., with $\langle\tilde{f}\rangle =\overline{f}$) has a
covariance function greater than or equal to the Fisher covariance:
\begin{equation} \label{cramer}
{\rm Cov}({\tilde{f}}) \succeq {\rm Cov_F}(f) .
\end{equation}
The Fisher covariance therefore characterises the minimum uncertainty of such
estimates.  For example, the quantity 
\[
(\Delta_F f_x)^2 := \left[{\rm Cov_F}(f)\right]_{xx} \]
represents 
a lower bound for the variance, $(\Delta\tilde{f}_x)^2$, of any unbiased
estimate of the displacement field at position $x$.
Roughly speaking, two ensembles differing by displacement field
$\overline{f}$ stand a decent chance of being distinguished 
at position $x$ if the Rayleigh-type criterion
$|\overline{f}_x|\geq \Delta_F f_x$ is satisfied.

To explicitly define the Fisher covariance function, let $F(f)$
denote the ``Fisher information matrix" of an ensemble described by
probability density functional $P[f]$ \cite{cover, cramer}, with
\begin{equation} \label{fishmat}
\left[ F(f)\right]_{x x'} := \int D\! f\, P \frac{\delta\ln P}{\delta f_x} 
\frac{\delta\ln P}{\delta f_{x'}}.
\end{equation}
Then ${\rm Cov_F}(f)$ is defined as the corresponding matrix inverse,
i.e.,
\begin{equation}  \label{finv}
{\rm Cov_F}(f)\ast F(f) = F(f) \ast {\rm Cov_F}(f) = \openone .
\end{equation}
Fisher information was introduced in 1925 for one-dimensional
statistical variables \cite{fish}, and Eq. (\ref{cramer}) 
is the continuous matrix version of the 
Cramer-Rao inequality \cite{cramer}, well known in classical statistics.
A particular case corresponds to the choice
$\tilde{f}=f$ in Eq. (\ref{cramer}), yielding
\begin{equation} \label{fcramer}
{\rm Cov}(f)\succeq{\rm Cov_F}(f) . 
\end{equation}
The Fisher covariance is therefore also a lower bound for the standard
covariance.  As this result is needed for establishing the connection
between Heisenberg and exact uncertainty relations in Sec. II.D, it
is useful to give a direct proof here.

First, let $A[f]$ denote an arbitrary functional of $f$ with $\langle
A\rangle < \infty$, and define
\begin{eqnarray} \label{mdef}
M_{xx'} & := &\left\langle\left(\frac{\delta\ln P}{\delta f_x} - \frac{\delta
A}{\delta f_x}\right) \left(\frac{\delta\ln P}{\delta f_{x'}}
-\frac{\delta A}{\delta f_{x'}}\right)\right\rangle\\ 
& = & \left[F(f)\right]_{xx'} +\left\langle\frac{\delta A}{\delta f_{x}}
\frac{\delta A}{\delta f_{x'}}\right\rangle - \int D\! f\,\left( \frac{\delta
P}{\delta f_x}\frac{\delta A}{\delta f_{x'}} + \frac{\delta P}{\delta
f_{x'}}\frac{\delta A}{\delta f_{x}}\right) \nonumber\\
& = & \left[F(f)\right]_{xx'} + \left\langle\frac{\delta A}{\delta
f_{x}}\frac{\delta A}{\delta f_{x'}} + 2\frac{\delta^2 A}{\delta f_x
\delta f_{x'}} \right\rangle .\nonumber
\end{eqnarray}
Restricting $A$ to have the quadratic form
\begin{equation} \label{adef}
A[f] = -\frac{1}{2} \int dxdx'\, K_{xx'}(f_x-h_x)(f_{x'}-h_{x'})
\end{equation}
for some symmetric matrix function $K$ and field $h$, and
noting that $M\succeq 0$ from Eq. (\ref{mdef}), then gives
\[
F(f) \succeq 2K - K\ast{\rm Cov}(f)\ast K , \]
where we have chosen $h=\langle f\rangle$ at the end of the calculation.  
Finally, Eq.
(\ref{fcramer}) follows by choosing $K$ to be the matrix inverse of
${\rm Cov}(f)$. 
Note that equality holds in Eq. (\ref{fcramer}) if and only
if $M\equiv 0$ in Eq. (\ref{mdef}), i.e., if and only if $\delta (\ln
P)/\delta f \equiv \delta A/\delta f$.  Thus this Cramer-Rao 
inequality is saturated
if and only if $P\sim\exp(A)$, i.e., noting Eq. (\ref{adef}), if and
only if the ensemble is Gaussian.

It is important to emphasise here that the Fisher
covariance function is a purely {\it classical} quantity, requiring no
reference to quantum theory whatsoever for its motivation or definition.  
However, it plays an important role as a measure of uncertainty
in an exact uncertainty relation for quantum fields, to be
derived in Sec. II.D.

\subsection{An exact uncertainty relation}

It has been shown that the momentum density of a bosonic quantum field
admits a natural decomposition into classical and nonclassical
components.  Moreover, Eqs. (\ref{comm}), (\ref{cond1}) and
(\ref{gdecomp}) imply that 
\[
[\hat{f}_x, \hat{g}^{cl}_{x'}] = 0,~~~~[\hat{f}_x,\hat{g}^{nc}_{x'}] =
i\hbar\delta(x-x') .\]
Hence it is the {\it nonclassical} component which is responsible for
the non-vanishing commutator in Eq. (\ref{comm}).  However, this
nonclassical component does not merely satisfy a Heisenberg-type inequality
analogous to Eq. (\ref{hfield}):  
it satisfies the {\it exact} uncertainty relation
\begin{equation} \label{geur}
{\rm Cov_F}(f) \ast {\rm Cov}(g^{nc}) = (\hbar^2/4) \openone .
\end{equation}
Thus, {\it the field uncertainty precisely determines the nonclassical
momentum uncertainty}, and vice versa.  

The exact uncertainty relation in Eq. (\ref{geur}) 
is the main result of this
Section.  It is remarkable that natural measures of uncertainty 
can be chosen which
precisely quantify the uncertainty principle for bosonic fields.
Note that the quantities appearing in Eq.~(\ref{geur}) above  are
operational in nature:  ${\rm Cov}_F(f)$ may be calculated from the
statistics of $\hat{f}$, and ${\rm
Cov}(g_{nc})$ may be calculated from the statistics of
$\hat{f}$ and $\hat{g}$, via Eq.~(\ref{covadd}).  
Thus the exact uncertainty relation directly relates the statistics
of the conjugate fields $\hat{f}$ and $\hat{g}$.

To demonstrate Eq. (\ref{geur}), note that for an ensemble of quantum
fields described by probability functional $P[f] = |\Psi[f]|^2$, the
Fisher information  follows from Eq. (\ref{fishmat}) as
\begin{eqnarray*}
\left[ F(f)\right]_{xx'} & = & 
\int D\! f\, \Psi^*\Psi\left[\Psi^{-1}\frac{\delta\Psi}{
\delta f_x} + c.c\right]\left[\Psi^{-1}\frac{\delta\Psi}{\delta f_{x'}} +
c.c.\right]\\
& = & \int D\! f \,\Psi^*\Psi\left[\Psi^{-1}\frac{\delta\Psi}{\delta f_x} -
c.c.\right]\left[\Psi^{-1}\frac{\delta\Psi}{\delta f_{x'}} -
c.c.\right]\\
&   & \mbox{} + 2\int D\! f\,\left[\frac{\delta\Psi^*}{\delta f_x}
\frac{\delta\Psi}{\delta f_{x'}} + c.c.\right]\\
& = & -(4/\hbar^2)\langle\hat{g}^{cl}_x\hat{g}^{cl}_{x'}\rangle +
(4/\hbar^2)\langle \hat{g}_x\hat{g}_{x'}\rangle,
\end{eqnarray*}
where the last line follows via Eqs. (\ref{srep}) and (\ref{gcl}).  Hence,
using Eqs. (\ref{gav}) and (\ref{covadd}),
\begin{equation} \label{fishnc}
{\rm Cov}(g^{nc}) = (\hbar^2/4) F(f) , 
\end{equation}
i.e., {\it the covariance of the nonclassical momentum density is proportional
to the Fisher information of the field}.  Eqs. (\ref{geur}) and
(\ref{fishnc}) are equivalent. 

The exact uncertainty relation in Eq.~(\ref{geur}), being a strict
equality for all states of the field, is clearly far
stronger in character than the corresponding Heisenberg-type relation in 
Eq.~(\ref{hfield}).  Moreover, it immediately implies the latter, since ${\rm
Cov}(g) \succeq {\rm Cov}(g^{nc})$ and ${\rm Cov}(f) \succeq {\rm
Cov_F}(f)$ from Eqs. (\ref{covadd}) and (\ref{fcramer}) respectively. 

Various applications of such exact 
uncertainty relations may be made, analogous
to those made in Refs. \cite{hallfish, eur} for quantum particles,
but will not be pursued here.  However, the notion that the momentum
density of a bosonic field 
decomposes into a classical and a nonclassical part, with the
uncertainty of the latter determined by the uncertainty of the field,
underlies the formulation of a new approach to quantum field
theory in Sec. III below.

\subsection{Multicomponent and complex bosonic fields}
It is of interest to briefly outline the form of exact uncertainty
relations for more general bosonic fields.  
Thus, let $f^a$
denote a multicomponent bosonic field with conjugate momentum density
$g^a$.  For example, $f^a$ may be a complex Klein-Gordon field $\phi$,
or the electromagnetic field $A^\mu$. This notation 
also covers the case where the index $a$ labels 
several interacting bosonic fields.
 
The non-vanishing equal-time commutation relation is now
\begin{equation}
\left[\hat{f}^a_x, \hat{g}^b_{x'}\right] = i\hbar\delta^{ab}\delta(x-x'),
\end{equation}
and the momentum density operator decomposes into classical and
nonclassical components as before, with
\begin{equation}
\hat{g}^a = \hat{g}^{cl, a} + \hat{g}^{nc, a} ,
\end{equation}
\begin{equation}
g^{cl, a}[f] = \frac{\hbar}{2i}\left[\Psi^{-1}\frac{\delta\Psi}{\delta
f^a} - (\Psi^*)^{-1}\frac{\delta\Psi^*}{\delta f^a}\right] .
\end{equation}
Clearly the components of $\hat{g}^{cl}$ commute with each other;  it
may also be checked that the components of $\hat{g}^{nc}$ commute (which
is necessary for the covariance function ${\rm Cov}(g^{nc})$ 
to be well defined).

The generalised covariance and Fisher information matrices are defined
by 
\begin{equation}
\left[{\rm Cov}(h)\right]^{ab}_{xx'} := \langle~(h^a_x)^*h^b_{x'}~\rangle
- \langle(h^a_x)^*\rangle \langle h^b_{x'}\rangle ,
\end{equation}
\begin{equation}
\left[ F(f)\right]^{ab}_{xx'} :=
\int D\! f\, P\frac{\delta\ln P}{\delta(f^a_x)^*}
\frac{\delta\ln P}{\delta f^b_{x'}} , 
\end{equation}
in analogy to Eqs. (\ref{hcov}) and (\ref{fishmat}) respectively, and
similarly are purely classical statistical quantities.

Finally, the corresponding exact uncertainty relation has precisely the
form of Eq. (\ref{geur}), providing that the definitions of the 
multiplicative identity and matrix multiplication are replaced by the
natural generalisations 
\[
\openone^{ab}_{xx'} := \delta^{ab}\delta(x-x') ,\]
\[
\left[ A\ast B\right]^{ab}_{xx'} := \sum_c\int dx''\, A^{ac}_{xx''}
B^{cb}_{x''x'} .\]
respectively.
\newpage
\section{Quantisation of fields from an exact uncertainty principle}

It has been shown that the momentum density of a
bosonic quantum field splits naturally and neatly into the sum of {\it
classical} and {\it nonclassical} parts, as per Eq. (\ref{gdecomp}).
Moreover, the uncertainty of the nonclassical part is {\it precisely
linked} to the uncertainty of the field itself, as per Eq. (\ref{geur}).
These results not only provide a mathematical link between the
statistics of conjugate quantum fields:  they suggest the existence of a
new bridge connecting classical field theory to quantum field theory - an
exact uncertainty principle.

Indeed, as demonstrated in this section, the assumption that an ensemble
of classical fields is subjected to ``nonclassical'' momentum
fluctuations, of a magnitude determined by the field uncertainty, leads
from the classical equations of motion to the equation of motion  of a
bosonic quantum field.  This result is restricted to the case of
field Hamiltonians quadratic in the momentum density, but nevertheless 
covers most classical fields of physical interest, including the
Klein-Gordon, electromagnetic and gravitational fields.

\subsection{Classical ensembles}

We first consider a real multicomponent classical field $f\equiv(f^a)$ 
with conjugate
momentum density $g\equiv(g^a)$, described by some Hamiltonian functional
$H[f,g,t]$.
Spatial coordinates are denoted by $x$ (irrespective of dimension), and
the values of field components $f^a$ and $g^b$ at position $x$ are
denoted by $f^a_x$ and $g^b_x$ respectively.

The equations of motion for an {\it ensemble} of such fields are
given by the Hamilton-Jacobi equation
\begin{equation} \label{mhj}
\frac{\partial S}{\partial t} + H[f,\delta S/\delta f, t] = 0 ,
\end{equation}
and the continuity equation
\begin{equation} \label{mcont}
\frac{\partial P}{\partial t} + \sum_a\int dx\, \frac{\delta}{\delta
f^a_x}\left(P\left.\frac{\delta H}{\delta g^a_x}\right|_{g=\delta
S/\delta f}\right) = 0 ,
\end{equation}
for the momentum potential $S$ and the probability density $P$,
as is reviewed in Appendix B.  These equations specify the motion of
the ensemble completely, where the momentum density associated with $f$
is given by $g=\delta S/\delta f$.

Defining the {\it ensemble Hamiltonian} $\tilde{H}$
as the functional integral
\begin{equation} \label{mham}
\tilde{H}[P, S, t]:= \langle H\rangle = \int D\!f\, PH[f, \delta
S/\delta f, t] ,
\end{equation}
the equations of motion for the dynamical variables 
$P$ and $S$ may be written in the
Hamiltonian form
\begin{equation} \label{mconj}
\frac{\partial P}{\partial t} = \frac{\delta\tilde{H}}{\delta S}, ~~~~~
\frac{\partial S}{\partial t} = - \frac{\delta\tilde{H}}{\delta P} .
\end{equation}
The variational derivative of a functional integral such as $\tilde{H}$,
with respect to a functional such as $P$ or $S$, is
discussed in Appendix A, and
the equivalence of Eqs. (\ref{mhj})
and (\ref{mcont}) to Eqs. (\ref{mconj}) follows directly from Eq.
(\ref{mform}).
Hence, in analogy to Eqs. (\ref{conj})
of Appendix B, $P$ {\it and} $S$ {\it may be regarded as 
canonically conjugate functionals, 
which evolve under the ensemble Hamiltonian} $\tilde{H}$ \cite{footact}.  Note 
from Eq. (\ref{mham}) that $\tilde{H}$ typically corresponds to the mean
energy of the ensemble.   

In what follows, we will only consider classical fields with 
Hamiltonian functionals quadratic in the momentum field
density, i.e., of the form
\begin{equation} \label{hquad}
H[f, g, t] = \sum_{a,b}\int dx\, K^{ab}_{x}[f] g^a_xg^b_x + V[f] .
\end{equation}
Here $K^{ab}_x[f]=K^{ba}_x[f]$ is a kinetic factor coupling components
of the momentum density, and $V[f]$ is some potential energy functional. 
The corresponding
{\it ensemble} Hamiltonian is given by Eq. (\ref{mham}).  Note that cross terms
of the form $g^a_xg^b_{x'}$ with $x\neq x'$
are not permitted in local field theories, and
hence are not considered here. 

\subsection{Momentum fluctuations $\Rightarrow$ quantum ensembles}

The ensemble Hamiltonian $\tilde{H}[P,S,t]$ in Eq. (\ref{mham}) is our classical
starting point for describing an ensemble of fields.  This starting
point must be modified in some way if one is to obtain new equations of
motion, to be identifed as describing a {\it quantum} 
ensemble of fields.  For example, the standard approach (canonical
quantisation) assumes the
existence of a complex field functional $\Psi[f,t]$ living in some Hilbert
space, and one simply writes down the
Schr\"{o}dinger equation $i\hbar\partial\Psi/\partial t = 
H[f,-i\hbar(\delta/\delta f),t]\Psi$ \cite{schweber}.  While this approach can
generally be made to work successfully, it provides {\it no} explanation for
the appearance of the nonclassical objects 
$\Psi$ and $\hbar$, a Hilbert space, and a linear operator equation.  Moreover,
for cases in which $K^{ab}_x$ in Eq. (\ref{hquad}) is dependent on $f$
(eg, gravitational fields),
the canonical approach is ambiguous as to the ordering of the functional
derivative operator.  

In contrast, we take a very different approach, where the only
{\it a priori} nonclassical feature is an 
assumption that the classical ensemble Hamiltonian $\tilde{H}$ must be
modified to 
take into account the existence of nonclassical fluctuations of the momentum
density, with the magnitude of the fluctuations determined by the
uncertainty in the field. This ``exact uncertainty'' approach is
motivated by the fact that bosonic fields satisfy exact uncertainty
relations such as Eq. (\ref{geur}) of Sec.~II. It leads to equations of
motion equivalent to the Schr\"{o}dinger equation above, with the added
advantage of a {\it unique} operator ordering.

Suppose then that $\delta S/\delta f$ is in fact an {\it average}
momentum density associated with field $f$, in the sense that the true momentum
density is given by 
\begin{equation} \label{noise}
g = \delta S/\delta f + N,
\end{equation}
where $N$ is a fluctuation field that vanishes on the average for any
given field $f$. Thus the physical meaning of $S$ changes to being an
{\it average} momentum potential. 
No specific underlying model for $N$ is assumed or necessary: in the approach
to be followed, one may in fact
interpret the ``source'' of the fluctuations as the field
uncertainty itself. Thus the nature of the fluctuation field is not
important -
its main effect is to remove any deterministic connection between $f$
and $g$. 

Since the momentum fluctuations may conceivably depend on the field $f$,
the average over such fluctuations for a given quantity $A[f,N]$ will be
denoted by $\overline{A}[f]$, and the average over fluctuations {\it
and} the field by $\langle A\rangle$.  Thus $\overline{N}\equiv 0$ 
by assumption, and in general $\langle A\rangle = \int
D\!f\,P[f] \overline{A}[f]$.  Assuming a quadratic dependence on
momentum density as per Eq. (\ref{hquad}), it follows 
that when the fluctuations are
significant the classical 
ensemble Hamiltonian $\tilde{H} = \langle H\rangle$ in Eq.~(\ref{mham}) should
be replaced by
\begin{eqnarray}
\tilde{H}' & = & \langle~ H[f, \delta S/\delta f + N, t]~\rangle\nonumber\\\
& = & \sum_{a,b}\int D\!f\,\int dx\, P K^{ab}_x \overline{( \delta
S/\delta f^a_x +N^a_x) (\delta S/\delta f^b_x +N^b_x)} + \langle
V\rangle\nonumber\\ \label{hprime}
& = & \tilde{H} + \sum_{a,b}\int D\!f\, \int dx\, P K^{ab}_x \overline{
N^a_xN^b_x} .  
\end{eqnarray}
Thus the momentum fluctuations lead to an additional nonclassical term in the 
ensemble Hamiltonian, dependent on the covariance matrix ${\rm
Cov}_x(N)$ of the fluctuations at position $x$, where
\begin{equation} \label{covn}
\left[{\rm Cov}_x(N)\right]^{ab} := \overline{N^a_xN^b_x} .  
\end{equation}

The additional term in Eq. (\ref{hprime}) is uniquely specified, up to a
multiplicative constant, by the following four assumptions:

{\it (1) Causality:}  $\tilde{H}'$ is an ensemble Hamiltonian for the
canonically conjugate functionals $P$ and $S$, which yields causal equations
of motion. Thus no higher than first-order functional derivatives can appear in
the additional term, implying that 
\[
{\rm Cov}_x(N) = \alpha(P, \delta P/\delta f_x, S, \delta S/\delta
f_x, f_x, t) 
\]
for some symmetric matrix function $\alpha$.

{\it (2) Independence:}  If the system comprises two independent
non-interacting subsystems 1 and 2, with factorisable probability
density functional $P[f^{(1)},f^{(2)}] = P_1[f^{(1)}] P_2[f^{(2)}]$, 
then the dependence of the subsystem fluctuations on $P$ only enters via
the corresponding probability densities $P_1$ and $P_2$ respectively. Thus
\[
\left.{\rm Cov}_x(N^{(1)})\right|_{P_1P_2} = \left.{\rm
Cov}_x(N^{(1)})\right|_{P_1}, ~~~~\left.{\rm
Cov}_x(N^{(2)})\right|_{P_1P_2} =\left.{\rm Cov}_x(N^{(2)})\right|_{P_2}
\]
for such a system.  
Note that this assumption implies that ensemble Hamiltonians are 
additive for independent non-interacting ensembles (as are the
corresponding actions \cite{footact}). 

{\it (3) Invariance:} The additional term transforms correctly under
linear canonical transformations of the field components.  Thus, noting 
that $f\rightarrow\Lambda^{-1}f$, $g\rightarrow\Lambda^Tg$ is a
canonical transformation for any invertible matrix $\Lambda$ (with
transpose $\Lambda^T$ and  
coefficients $\Lambda_{ab}$), which preserves the quadratic form of $H$
in Eq. (\ref{hquad}) and leaves the momentum potential $S$ invariant (since
$\delta/\delta f\rightarrow \Lambda^T\delta/\delta f$), one has from 
Eq. (\ref{noise}) that
$N\rightarrow\Lambda^TN$, and hence that
\[ {\rm Cov}_x(N) \rightarrow \Lambda^T {\rm Cov}_x(N)\Lambda ~~~{\rm for}
~~~f\rightarrow\Lambda^{-1}f .\]
Note that for single-component fields this reduces to a scaling relation
for the variance of the fluctuations at each point $x$.
  
{\it (4) Exact uncertainty:} The uncertainty of the momentum
fluctuations at any given position and time, as characterised by ${\rm
Cov}_x(N)$, is specified by the field uncertainty at that position and
time.  Thus, since the field uncertainty is completely determined by the
probability density functional $P$, it follows that ${\rm Cov}_x(N)$
cannot depend on $S$, nor explicitly on $t$. 

It is seen that the first three assumptions are essentially 
classical in nature, 
while the fourth assumption is not: it postulates an exact connection
between the nonclassical momentum uncertainty and the field uncertainty.   
Remarkably, these assumptions lead
directly to equations of motion of a bosonic quantum field, as shown by
the following Theorem and Corollary (proofs are given in Appendix C).

{\bf Theorem:}  {\it The above assumptions of causality, independence,
invariance, and 
exact uncertainty imply that} 
\begin{equation} \label{theorem}
\overline{N^a_xN^b_x} = C (\delta P/\delta f^a_x) (\delta P/\delta
f^b_x)/P^2 ,
\end{equation}
{\it where $C$ is a positive universal constant.}  

The theorem thus yields a unique form for the additional term in Eq.
(\ref{hprime}), up to a multiplicative constant $C$.  The classical
equations of motion for the ensemble are recovered in the limit of small
fluctuations, i.e., in the limit
$C\rightarrow0$.  Note that one cannot make the identification 
$N^a_x\sim(\delta P/\delta f^a_x)/P$ from Eq. (\ref{theorem}),
as this is inconsistent with the fundamental property
$\overline{N^a_x}=0$.

The main result of this Section is the following
corollary (proved in Appendix C):

{\bf Corollary:} {\it The equations of motion corresponding to the
ensemble Hamiltonian $\tilde{H}'$ can be expressed as the single complex
equation}
\begin{equation} \label{corollary}
i\hbar\frac{\partial\Psi}{\partial t} =
H[f,-i\hbar\delta/\delta f, t] \Psi = -\hbar^2 \left(\sum_{a,b}\int dx\,
\frac{\delta}{\delta f^a_x} K^{ab}_x[f] \frac{\delta}{\delta f^b_x}
\right) \Psi +
V[f]\Psi ,
\end{equation}
{\it where one defines}
\begin{equation} \label{hdef}
\hbar := 2\sqrt{C},~~~~~~\Psi := \sqrt{P} e^{iS/\hbar} .
\end{equation}

Eq. (\ref{corollary}) may be recognised as the Schr\"{o}dinger equation
for a quantum bosonic field, and hence the goal of
deriving this equation, via an exact uncertainty principle for 
nonclassical
momentum fluctuations acting on a classical ensemble, has been achieved.  
The ensemble of fields corresponding to ensemble Hamiltonian
$\tilde{H}'$ will therefore be called the {\it quantum ensemble}
corresponding to $\tilde{H}$. It is remarkable that the four assumptions
of causality, independence, invariance and exact uncertainty lead to a {\it
linear operator} equation. 

Note that the
exact uncertainty approach specifies a {\it unique} operator ordering,
$(\delta/\delta f^a_x)K^{ab}_x(\delta/\delta f^b_x)$,
for the functional derivative operators in Eq. (\ref{corollary}).  Thus
there is no ambiguity in the ordering for cases where $K^{ab}_x$ depends on
the field $f$, in contrast to traditional
approaches (eg, the Wheeler-deWitt equation, discussed in Sec.
III.D below).  The above results generalise straightforwardly to complex
classical fields.  

Finally, it may be remarked that the equations of motion of a classical
ensemble may be subject to some imposed constraint(s) on $f$, $P$, $S$ and
$\tilde{H}$.  For example, each
member of an ensemble of electromagnetic fields may have the Lorentz gauge 
imposed (see Sec III.C
below). As a guiding principle, we will require that the corresponding
{\it quantum} ensemble is subject to the same constraint(s) on $f$, $P$, $S$
and $\tilde{H}'$.  This will ensure a meaningful classical-quantum
correspondence for the results of field measurements.  However,
consistency of the quantum equations of motion with a given set of
constraints is not guaranteed by the above Theorem and Corollary, and so
must be checked independently for each case.

\subsection{Example: Electromagnetic field}

The electromagnetic field is described, up to gauge invariance, 
by a 4-component field
$A^\mu$.  In the Lorentz gauge all physical fields satisfy 
$\partial_\mu A^\mu\equiv 0$, and the classical 
equations of motion in vacuum are
given by $\partial^\nu\partial_\nu A^\mu=0$.  These follow,
for example, from the Hamiltonian \cite{schweber} 
\begin{equation} \label{emham}
H_{GB}[A,\pi] =
(1/2)\int dx\, \eta_{\mu\nu}\left(\pi^\mu\pi^\nu - \nabla A^\mu\cdot\nabla
A^\nu\right) ,
\end{equation}
where $\eta_{\mu\nu}$ denotes the Minkowski tensor, $\pi^\mu$ denotes
the conjugate momentum density, and $\nabla$ denotes the spatial
derivative. Here $H_{GB}$ corresponds to the gauge-breaking 
Lagrangian $L=-(1/2)\int dx\,A^{\mu,\nu}A_{\mu,\nu}$, and is seen 
to have the quadratic form of Eq.~(\ref{hquad}) 
(with $K^{\mu\nu}_x=\eta_{\mu\nu}/2$).  Hence the exact uncertainty  
uncertainty approach immediately implies that 
a {\it quantum} ensemble of electromagnetic fields obeys the Schr\"{o}dinger
equation
\begin{equation}
i\hbar(\partial\Psi/\partial t) = H_{GB}[A,-i\hbar(\delta/\delta A)]
\Psi ,
\end{equation} 
in agreement with the Gupta-Bleuler formalism \cite{schweber}.

Note that the probability of a member of the classical 
ensemble not satisfying the
Lorentz gauge condition $\partial_\mu A^\mu\equiv 0$ is zero by
assumption, i.e., the Lorentz gauge is equivalent to the condition
that the product $(\partial_\mu A^\mu) P[A]$ vanishes for all
physical fields.  For the quantum ensemble to satisfy this 
condition, as per the guiding principle discussed at the end of
Section III.B above, one equivalently requires,
noting Eq. (\ref{hdef}), that 
\[
(\partial_\mu A^\mu) \Psi[A] =0 .\]
As is well known, this constraint, if initially satisfied, is satisfied
for all times \cite{dirac} (as is the alternative weaker constraint that
only the 4-divergence of the positive frequency part of the field vanishes
\cite{schweber}). Hence the evolution of the quantum
ensemble is consistent with the Lorentz gauge.
It would be of interest to derive the consistency of this constraint
directly from the equations of motion, Eqs. (\ref{modcont}) and
(\ref{modhj}), for $P$ and $S$.
 
Finally, it is well known that one  can 
also obtain the classical equations of motion via an
alternative Hamiltonian, by exploiting the degree of freedom left by the
Lorentz gauge to remove a dynamical 
coordinate (corresponding to the longitudinal polarisation).  In
particular, since $\partial_\mu A^\mu$ is invariant under 
$A^\mu\rightarrow A^\mu +\partial^\mu \chi$ for any function $\chi$
satisfying $\partial^\nu\partial_\nu\chi=0$, one may completely
fix the gauge in a given Lorentz frame by choosing $\chi$ such that
$A^0=0$.  One thus obtains, writing $A^\mu\equiv(A^0,{\bf A})$, the
radiation gauge $A^0=0$, $\nabla\cdot{\bf A}=0$.  The classical
equations of motion for
${\bf A}$ ($\partial^\nu\partial_\nu{\bf A}=0$), follow, for example,
from the Hamiltonian 
\begin{equation} \label{hc}
H_R[{\bf A},{\bf E}] = (1/2)\int dx\, ({\bf E}\cdot{\bf E} +
|\nabla\times{\bf A}|^2) ,
\end{equation}
where ${\bf E}$ denotes the conjugate momentum density. Here $H_R$
corresponds to the standard Lagrangian $L=-(1/4)\int dx\,F^{\mu\nu}F_{\mu\nu}$,
with  $A^0\equiv 0$.  This Hamiltonian once again has the quadratic
form of Eq. (\ref{hquad}), and hence the exact uncertainty approach 
yields the corresponding Schr\"{o}dinger equation
\begin{equation}
i\hbar(\partial\Psi/\partial t) = H_R[{\bf A},-i\hbar(\delta/\delta {\bf
A})]\Psi
\end{equation} 
for a quantum ensemble of electromagnetic 
fields in the radiation gauge  (this is also
the form of the Schr\"{o}dinger equation obtained via the
Schwinger-Tomonaga formalism \cite{wheeler}).  Note that 
the momentum
density ${\bf E}$ is in fact the electric field, and hence, in this
case, the exact uncertainty approach corresponds to adding nonclassical
fluctuations to the electric field components, with the fluctuation uncertainty
determined by the uncertainty in the vector potential ${\bf A}$.

\subsection{Example: Gravitational field}

The gravitational field is described, up to arbitrary coordinate
transformations, by the metric tensor $g\equiv(g_{\mu\nu})$.  The corresponding
invariant length may be decomposed as \cite{dw}
\[
ds^2 = g_{\mu\nu}dx^\mu dx^\nu  = - (\alpha^2-{\bm \beta}\cdot {\bm \beta})dt^2
+ 2\beta_idx^idt+\gamma_{ij}dx^idx^j , \]
in terms of the lapse function $\alpha$, the shift function ${\bm
\beta}$, and the spatial 3-metric $\gamma\equiv(\gamma_{ij})$.  The
equations of motion are the Einstein field equations, which follow from
the Hamiltonian functional \cite{dw}
\begin{equation} \label{hdw}
H[\gamma, \pi,\alpha,{\bm \beta}] = \int dx\, \alpha{\cal H}_G[\gamma, \pi] -
2\int dx\,\beta_i\pi^{ij}_{~~|j}~ ,
\end{equation}
where $\pi\equiv(\pi^{ij})$ denotes the momentum density conjugate to
$\gamma$, $|j$ denotes the covariant 3-derivative, and the Hamiltonian
density ${\cal H}_G$ is given by
\begin{equation} \label{hg}
{\cal H}_G = 
(1/2) G_{ijkl}[\gamma]\pi^{ij}\pi^{kl} - 2\,
{}^{(3)}\!R[\gamma](\det\gamma)^{1/2} .
\end{equation}
Here ${}^{(3)}\!R$ is the curvature scalar corresponding to
$\gamma_{ij}$, and
\[
G_{ijkl}[\gamma] =(\gamma_{ik}\gamma_{jl}+\gamma_{il}\gamma_{jk}
-\gamma_{ij}\gamma_{kl})(\det \gamma)^{-1/2} . \]

The Hamiltonian functional $H$ corresponds to the standard Lagrangian 
$L=\int dx\,(-\det
g)^{1/2} R[g]$, 
where the momenta $\pi^0$ and $\pi^i$ conjugate to $\alpha$ and $\beta_i$
respectively vanish identically \cite{dw}.  However, the lack of dependence of
$H$ on $\pi^0$ and $\pi^i$ is consistently maintained only if the rates
of change of these momenta also vanish, i.e., noting Eq. (\ref{conj}) of
Appendix B, only if the constraints \cite{dw}
\begin{equation} \label{constraint}
\delta H/\delta\alpha = {\cal H}_G = 0,~~~~\delta H/\delta\beta_i = -2
\pi^{ij}_{~~| j}=0
\end{equation}
are satisfied.  Thus the dynamics of the field
is independent of $\alpha$ and ${\bm \beta}$, so that these 
functions may be fixed arbitrarily \cite{dw}.  Moreover, these
constraints 
immediately yield $H=0$ in Eq. (\ref{hdw}), and hence 
the system is static, with no explicit time dependence \cite{dw}.

It follows that in the Hamilton-Jacobi formulation of the 
equations of motion (see Appendix B), 
the momentum potential $S$ is 
independent of $\alpha$, ${\bm \beta}$ and $t$.  Noting that
$\pi\equiv\delta S/\delta\gamma$ in this formulation, Eqs.
(\ref{constraint}) therefore yield the corresponding constraints
\begin{equation} \label{cons}
\frac{\delta S}{\delta \alpha} = \frac{\delta S}{\delta\beta_i} = 
\frac{\partial S}{\partial t} = 0,~~~~
\left(\frac{\delta S}{\delta\gamma_{ij}}\right)_{| j}=0  
\end{equation} 
for $S$.  As shown by Peres \cite{peres}, a given functional $F[\gamma]$ of the
3-metric is invariant under spatial coordinate transformations if and
only if $(\delta 
F/\delta\gamma_{ij})_{|j}=0$, and hence the fourth constraint
in Eq.~(\ref{cons}) is equivalent to the invariance of $S$ under such
transformations.  This fourth constraint moreover implies that the
second 
term in Eq.~({\ref{hdw}) may be dropped from the Hamiltonian, 
yielding the 
reduced Hamiltonian 
\begin{equation} \label{hgr}
H_G[\gamma,\pi,\alpha] = \int dx\,\alpha {\cal H}_G[\gamma,\pi] 
\end{equation}
in the Hamiltonian-Jacobi formulation \cite{peres, gerlach}.

For an {\it ensemble} of classical gravitational fields, the 
independence of the dynamics with respect to $\alpha$, ${\bm \beta}$ and 
$t$ implies that members of the ensemble are distinguishable only by their
corresponding 3-metric $\gamma$.  Moreover, it 
is natural to impose the additional geometric requirement that the ensemble is
invariant under spatial coordinate transformations.  One therefore has
the constraints
\begin{equation} \label{conp}
\frac{\delta P}{\delta\alpha}= \frac{\delta P}{\delta\beta_i} =
\frac{\partial P}{\partial t} = 0,~~~~
\left(\frac{\delta P}{\delta\gamma_{ij}}\right)_{| j}=0 
\end{equation}
for the corresponding probability density functional $P[\gamma]$,
analogous to Eq. (\ref{cons}). The first two constraints imply that
ensemble averages only involve integration over $\gamma$.
 
Noting Eq.~(\ref{hg}), the Hamiltonian $H_G$ in Eq.~(\ref{hgr}) 
has the quadratic form of Eq. (\ref{hquad}). Hence 
the exact uncertainty approach is applicable, and immediately leads to the
Schr\"{o}dinger equation 
\begin{equation} \label{seq}
i\hbar\partial\Psi/\partial t=\int dx\,\alpha {\cal H}_G[\gamma,
-i\hbar(\delta/\delta\gamma)]\Psi 
\end{equation}
for a {\it quantum} ensemble of gravitational fields, as per the
Corollary of Sec.~III.B.  

As discussed at the end of Sec.~III.B, we follow the guiding
principle that all constraints imposed on the classical ensemble should
be carried over to corresponding constraints on the quantum ensemble.  
Thus, from Eqs.~(\ref{cons}) and (\ref{conp}) we require that $P$ and
$S$, and hence $\Psi$ in Eq.~(\ref{hdef}), are independent of $\alpha$,
${\bm \beta}$ and $t$ and invariant under spatial coordinate
transformations, i.e., 
\begin{equation} \label{conpsi}
\frac{\delta \Psi}{\delta\alpha}= \frac{\delta \Psi}{\delta\beta_i} =
\frac{\partial \Psi}{\partial t} = 0,~~~~
\left(\frac{\delta \Psi}{\delta\gamma_{ij}}\right)_{| j}=0 .
\end{equation}
Applying the first and third of these constraints to Eq.~(\ref{seq})
immediately yields, via Eq.~(\ref{hg}),  
the reduced Schr\"{o}dinger equation
\begin{equation} \label{wdw}
{\cal H}_G[\gamma, -i\hbar(\delta/\delta \gamma)]\Psi = (-\hbar^2/2)
\frac{\delta}{\delta \gamma_{ij}}G_{ijkl}[\gamma]\frac{\delta
}{\delta \gamma_{kl}}\Psi - 2\,{}^{(3)}\!R[\gamma](\det\gamma)^{1/2}
\Psi = 0 ,
\end{equation} 
which may be recognised as  
the Wheeler-deWitt equation in the metric representation \cite{dw}. 
The consistency of Eqs.
(\ref{conpsi}) and (\ref{wdw}) is well known \cite{dw}. 

A notable feature of Eq.~(\ref{wdw}) is that the Wheeler-deWitt equation
has not only been derived from an exact uncertainty principle: it has, as 
a consequence of Eq.~(\ref{corollary}),
been derived with a {\it precisely}
defined operator ordering (with $G_{ijkl}$ sandwiched between the two
functional derivatives).
Thus the exact uncertainty approach does not admit
any ambiguity in the description of quantum gravity, unlike the standard
approach \cite{dw}.  Such removal of ambiguity is essential to
making definite physical predictions, and hence may be
regarded as an advantage of the exact uncertainty approach. 

For example, Kontoleon and Wiltshire \cite{wiltshire} have pointed
out that Vilenkin's prediction of inflation in minisuperspace,
from a corresponding Wheeler-deWitt equation with
``tunneling'' boundary conditions \cite{vilenkin},
depends critically upon the operator ordering used.
In particular, considering the class of orderings defined by an integer
power $p$, with corresponding
Wheeler-deWitt equation \cite{vilenkin}
\begin{equation} \label{vile}
\left[\frac{\partial^2}{\partial a^2} +
\frac{p}{a}\frac{\partial}{\partial a} - \frac{1}{a^2}\frac{\partial^2}{
\partial \phi^2} - U(a,\phi)\right]\Psi = 0 
\end{equation}
(for a Friedmann-Robertson-Walker metric coupled to a scalar field $\phi$), 
Kontoleon and Wiltshire show that Vilenkin's approach fails for orderings
with $p\geq 1$ \cite{wiltshire}.  Moreover, they suggest
that the only natural ordering is in fact the ``Laplacian'' ordering
corresponding to $p=1$, which has been justified on geometric grounds by
Hawking and Page \cite{hawking}.  

However, noting that the relevant Hamiltonian functional in Eq.~(2.7) of
Ref.~\cite{vilenkin} is quadratic in the momentum densities of the
metric and the scalar field, the exact uncertainty approach may be
applied, and yields the
Wheeler-deWitt equation corresponding to $p=-1$ in Eq. (\ref{vile}). 
Hence the criticism in Ref. \cite{wiltshire} is avoided.  
One also has the nice feature
that  the associated Wheeler-deWitt equation
can be exactly solved for this ``exact uncertainty'' ordering \cite{vilenkin}.

\section{Discussion}

The two main results of this paper are (i) an exact uncertainty relation,
Eq. (\ref{geur}), valid for all bosonic quantum fields, and (ii) the derivation 
of the
quantum equation of motion, Eq. (\ref{corollary}), from an exact
uncertainty principle, for fields with
Hamiltonian functionals quadratic in the momentum density.
These results are conceptually connected by the
notion of the momentum density splitting into the sum of classical and
nonclassical components, with the uncertainty of the nonclassical
component determined by the uncertainty of the field. Moreover, they 
generalise similar results obtained in Refs. 
\cite{hallfish, eur,hallreg, bamberg} for quantum particles.  

It is important to emphasise that the exact uncertainty approach in
Sec.~III.B 
does {\it not} assume the existence of a complex amplitude functional
$\Psi[f]$, nor the representation of fields by operators, nor the
existence of a universal constant $\hbar$ with units of action, nor the
existence of a linear operator equation in some Hilbert space.  
Only the assumptions of
causality, independence, invariance and exact uncertainty are required, all
formulated in terms of a {\it single}
nonclassical element (the uncertainty 
introduced by the momentum fluctuation $N$).  Since uncertainty is at
the conceptual core of quantum mechanics, this is an elegant and
pleasing result.

The assumptions used also provide an intuitive picture for the origin of the
Schr\"{o}dinger equation for bosonic fields, 
as arising from nonclassical fluctuations of
the momentum density.  Of course this picture has limitations - the
fluctuations essentially arise from the uncertainty of the field itself, rather
than from some external source, and hence are 
most certainly ``nonclassical'' rather than ``classical'' 
in nature.  

A minimalist interpretation of the exact uncertainty approach, based on
Eqs. (\ref{noise}) and (\ref{theorem}),
is that every {\it physical} field has an intrinsic uncertainty, 
the nature of which precludes
a deterministic relationship between the field and its
conjugate momentum density.  However, the {\it degree} of indeterminism in
this relationship {\it is} precisely quantifiable, in a statistical sense,
and is directly connected to the ensemble representation of the
inherent field uncertainty.

The exact uncertainty approach is very different to
the ``causal interpretation'' of Bohm and co-workers \cite{holland}.
In the latter it is assumed that there is a pre-existing complex
amplitude functional $\Psi[f]=\sqrt{P}\exp(iS/\hbar)$ 
obeying a Schr\"{o}dinger equation, 
which acts upon a single classical field  via
the addition of a ``quantum potential", $Q[P]$, to the classical Hamiltonian.
It is further assumed that the momentum density is precisely 
$g\equiv\delta S/\delta f$, and that physical ensembles of fields have 
probability density functional $P=|\Psi|^2$.
In contrast, the exact
uncertainty approach does not postulate the existence of adjunct
amplitudes and potentials, the Schr\"{o}dinger equation 
directly represents the evolution of an ensemble rather than of an 
external amplitude functional acting on individual systems 
(and is derived rather than postulated), 
and the basic tenet in Eq. (\ref{noise}) 
is that $g\neq\delta S/\delta f$.

It is of interest to consider the scope and limitations of the exact
uncertainty approach to physical systems.  Its applicability to
nonrelativistic quantum particles \cite{hallreg, bamberg, note}, and now
to bosonic quantum fields with Hamiltonians quadratic in the momentum
density (which include all relativistic integer-spin fields), 
has been demonstrated. 
It is also, indirectly, applicable
to the non-quadratic Hamiltonian functional
of a nonrelativistic boson field (corresponding to second quantisation
of the
usual nonrelativistic Schr\"{o}dinger equation), in the sense that this
field may be obtained as a low-energy limit of the complex Klein-Gordon
field
\cite{brown} (to which the exact uncertainty approach directly applies).

In this paper the basic Schr\"{o}dinger equation for bosonic fields
has been obtained, with the advantageous feature of a
unique operator ordering in cases where other approaches are ambiguous.   
It would be of interest to consider further issues, such as the
representation of general physical observables by operators 
(addressed for the case of particles in Ref. \cite{hallreg}), boundary
conditions, infinities, etc, from the new perspective on the conceptual
and technical basis of
quantisation offered by the exact uncertainty approach.

Finally, a major question to be addressed in the future is whether the exact
uncertainty approach is applicable to fermionic quantum fields. These have
two features which present challenges: the corresponding
ensemble Hamiltonian is usually linear in the momentum density, and the
anticommutation relations make it difficult to connect the equations of
motion with corresponding classical equations of motion in the limit as
$\hbar\rightarrow 0$.  One possible approach is to determine whether
exact uncertainty relations exist for such fields, analogous to Eq.
(\ref{geur}), as these would presumably suggest the statistical
properties required by suitable ``nonclassical'' fluctuations.

\appendix
\section{Functional derivatives and integrals}
Here necessary definitions and properties of functionals are noted,
including variational properties of functional integrals.
Since manipulations in the paper are of a purely formal character,
we avoid discussions of regularisations and discretizations needed to
fully define certain quantities.

A functional, $F[f]$, is a mapping from a set of physical fields
(assumed to form a vector space) to the real or complex numbers, and the
functional derivative of $F[f]$ is defined via the variation of $F$ with
respect to $f$, i.e., 
\begin{equation} \label{fderiv}
\delta F= 
F[f+\delta f] - F[f] = \int dx\, \frac{\delta F}{\delta f_x}\, \delta\!f_x
\end{equation}
for arbitrary infinitesimal variations $\delta f$.  Thus the functional
derivative is a 
field density, $\delta F/\delta f$, having the value $\delta
F/\delta f_x$ at position $x$.  For curved spaces one may explicitly 
include a volume element in the integral, thus redefining the functional
derivative by a multiplicative function of $x$; however, this is 
merely a matter of taste
and will not be adopted here.  The functional derivative is
assumed to always exist for the functionals in this paper.

It follows directly from Eq.~(\ref{fderiv}) that the functional
derivative satisfies product and chain rules analogous to ordinary
differentiation. The choice $F[f] = f_{x'}$ in Eq.
(\ref{fderiv}) yields
$\delta f_{x'}/\delta f_x = \delta(x-x')$ as required in Eqs.
(\ref{comm}) and (\ref{srep}). 
Moreover, if the field depends on some parameter, $t$ say, then
choosing $\delta f_x = f_x(t+\delta t) - f_x(t)$ in Eq. (\ref{fderiv})
yields
\begin{equation} \label{rate}
\frac{dF}{dt} = \frac{\partial F}{\partial t} + \int dx\,\frac{\delta
F}{\delta f_x} \frac{\partial f_x}{\partial t} 
\end{equation}
for the rate of change of $F$ with respect to $t$.

Functional integrals correspond to integration  of functionals over the
vector space of physical fields (or equivalence classes thereof).  
The only property we require for this paper is the existence of a measure
$D\!f$ on this vector space which is {\it translation invariant}, i.e.,
$\int D\!f \equiv \int D\!f'$ for any translation $f' = f + h$ (which 
follows immediately, for example, from the discretisation approach to
functional integration \cite{brown}). In particular, this 
property implies the useful result 
\begin{equation} \label{div}
\int D\!f\,\frac{\delta F}{\delta f} = 0~~~{\rm for}~~~\int D\!f\,
F[f] < \infty ,
\end{equation}
which is used repeatedly below and in the text.  Eq. (\ref{div}) follows
by noting that the finiteness condition and translation invariance imply
\[ 0 = \int D\!f\, (F[f+\delta f] - F[f]) = \int dx\,\delta\!f_x \left(
\int D\!f\,
\delta F/\delta f_x\right) \]
for arbitrary infinitesimal translations.

Thus, for example, if $F[f]$ has a finite expectation value
with respect to some probability density functional $P[f]$, then 
Eq. (\ref{div}) yields the ``integration by parts'' formula
\[
\int D\!f\,P (\delta F/\delta f) = -\int D\!f\, (\delta P/\delta f)F .\]
Moreover, from Eq. (\ref{div}) the total probability,
$\int D\!f\, P$,
is conserved for any probability flow satisfying a continuity equation of
the form
\begin{equation} \label{gcont}
\frac{\partial P}{\partial t} + \int dx\, \frac{\delta}{\delta f_x} \left[ 
PV_x\right] = 0 ,
\end{equation}
providing that the average flow rate, $\langle V_x\rangle$, is finite. 

Finally, consider a functional integral of the form
\begin{equation} \label{mdensity}
I[F] = \int D\!f\,\xi(F, \delta F/\delta f) ,
\end{equation}
where $\xi$ denotes any function of some functional $F$ and its functional
derivative.  Variation of $I[F]$ with respect to $F$ then gives, to
first order,

\begin{eqnarray*}
\delta I = I[F+\delta F]-I[F] & = & \int D\!f \,\left\{
(\partial \xi/\partial F)\delta F +\int dx\, \left[
\partial \xi/\partial (\delta F/\delta f_x)\right]\left[\delta
(\delta F)/\delta f_x\right]\right\}\\
& = & \int D\!f\, \left\{(\partial \xi/\partial F) - \int dx\,
\frac{\delta}{\delta f_x} \left[ \partial \xi/\partial (\delta
F/\delta f_x)\right]\right\}\delta F\\
& & \mbox{} + \int dx\,\int D\!f\, \frac{\delta}{\delta f_x} \left\{
\left[ \partial \xi/\partial (\delta F/\delta f_x)\right]\delta F
\right\} .
\end{eqnarray*}
Assuming that the functional integral of the expression in curly
brackets in the last term is finite, this term vanishes from Eq.
(\ref{div}), yielding  the result
\[ \delta I = \int D\!f\, \frac{\delta I}{\delta F}\, \delta F \]
analogous to Eq.~(\ref{fderiv}), where the variational derivative
$\delta I/\delta F$ is defined by
\begin{equation} \label{mform}
\frac{\delta I}{\delta F} = \frac{\partial \xi}{\partial F}
- \int dx\,\frac{\delta}{\delta f_x}\left[\frac{\partial \xi}{
\partial (\delta F/\delta f_x)}\right] .
\end{equation}
A similar result holds for multicomponent fields, with summation over
the discrete index $a$ in the second term.

\section{ Hamilton-Jacobi field theory}
The salient aspects of the Hamilton-Jacobi formulation of classical
field theory \cite{goldstein}
are collected here, with particular attention to the origin
of the associated continuity equation for {\it ensembles} of classical
fields, required in Sec. III.

Two classical fields $f$, $g$ are canonically conjugate if there is a
Hamiltonian functional $H[f,g,t]$ such that 
\begin{equation} \label{conj}
\partial f/\partial t = \delta H/\delta g, ~~~~\partial g/\partial t =
-\delta H/\delta f .
\end{equation}
These equations follow from the action principle $\delta A = 0$, with
action functional $A = \int dt\,[-H + \int dx\,g_x(\partial f_x/\partial
t)]$.  The rate of change of an arbitrary functional
$G[f,g,t]$ follows from Eqs. (\ref{rate}) and (\ref{conj}) as
\[
\frac{dG}{dt} = \frac{\partial G}{\partial t} + \int dx\,\left(
\frac{\delta G}{\delta f_x} \frac{\delta H}{\delta g_x} - \frac{\delta
G}{\delta g_x}\frac{\delta H}{\delta f_x} \right)
=: \frac{\partial G}{\partial
t} + \{ G,H\} , 
\]
where $\{~,~\}$ is a generalised Poisson bracket.

A canonical transformation maps $f$, $g$ and $H$ to $f'$, $g'$ and $H'$,
such that the equations of motion for the latter retain the canonical
form of Eq. (\ref{conj}).  Equating the variations of the corresponding
actions $A$ and $A'$ to zero, it follows that all physical trajectories
must satisfy
\[
-H + \int dx\, g_x(\partial f_x/\partial t) = -H' + \int dx\,g_x'(\partial
f_x'/\partial t) + dF/dt \]
for some ``generating functional" $F$.
Now, any two of the fields $f,g, f', g'$ determine the remaining two fields for
a given canonical transformation.  Choosing $f$ and $g'$ as the two
independent fields, defining the new generating functional $G[f,g',t] =
F + \int dx\, f_x'g_x'$, and using Eq.~(\ref{rate}), then yields 
\[
H' = H + \frac{\partial G}{\partial t} + \int dx\, \left[ \frac{\partial
f_x}{\partial t}\left( \frac{\delta G}{\delta f_x} -g_x\right) +
\frac{\partial g_x'}{\partial t}\left(\frac{\delta G}{\delta g_x'} -
f_x'\right)\right] \]
for all physical trajectories.  The terms in round brackets 
therefore vanish identically, yielding the generating relations
\begin{equation} \label{sgen}
H' = H + \partial G/\partial t,~~~~ g=\delta G/\delta
f,~~~~f'=\delta G/\delta g' .
\end{equation}
A canonical transformation is thus completely specified by the associated
generating functional $G$.

To obtain the {\it Hamilton-Jacobi} formulation of the  equations of
motion, consider a
canonical transformation to fields $f'$, $g'$ which are time-independent
(eg, to the fields $f$ and $g$ at some fixed time $t_0$). From 
Eq.~(\ref{conj}) one may  
choose the corresponding Hamiltonian $H'\equiv 0$ 
without loss of generality, and hence from 
Eq.~(\ref{sgen}) the momentum density and the associated
generating functional $S$ are specified by the functional equations
\begin{equation} \label{hj}
g=\frac{\delta S}{\delta f},~~~~
\frac{\partial S}{\partial t} + H[f, \delta S/\delta f,t] = 0 .
\end{equation}
The latter is the desired Hamilton-Jacobi equation. Solving this equation for
$S$ is equivalent to solving Eqs.~(\ref{conj}) for $f$ and $g$. 

Note that along a physical trajectory one has $g'\equiv$ constant, 
and hence from Eqs. (\ref{rate}) and
(\ref{hj}) that  
\[
\frac{dS}{dt} = \frac{\partial S}{\partial t} + \int dx\, \frac{\delta
S}{\delta f_x} \frac{\partial f_x}{\partial t} = -H + \int
dx\,g_x\frac{\partial f_x}{\partial t} = \frac{dA}{dt} . \]
Thus the Hamilton-Jacobi functional $S$ is equal to the action
functional $A$, up to an additive 
constant.  This relation underlies the connection
 between the derivation of the Hamilton-Jacobi equation from a
particular type of canonical transformation, as above, and the
derivation from a particular type of variation of the action, as per the
Schwinger-Tomonaga formalism \cite{wheeler,schwinger}.  In the latter case the
time parameter $t$ is replaced by the multi-time parameter
$\sigma$. 

The Hamilton-Jacobi formulation has the interesting feature that once $S$
is specified, the momentum density is determined by the relation
$g=\delta S/\delta f$, i.e., it is a functional of $f$.  Thus, unlike
the Hamiltonian formulation of Eqs. (\ref{conj}), an {\it ensemble} of
fields is specified by a probability density functional $P[f]$, not by a
phase space density functional $P[f,g]$. 
In both cases, the equation of motion for $P$ corresponds to the
conservation of probability, i.e., to a continuity equation as per Eq.
(\ref{gcont}).  In particular, since in the Hamilton-Jacobi formulation
the rate of change of the field $f$ follows from Eqs. (\ref{conj})
and (\ref{hj}) as the functional 
\[
V_x[f] = \partial f_x/\partial t = (\delta H/\delta g_x)
\left|{}_{g=\delta S/\delta f}\right. ,\]
the associated continuity equation for an ensemble of fields
follows from Eq. (\ref{gcont}) as  \cite{footnoteb}
\begin{equation} \label{hjcont}
\frac{\partial P}{\partial t} + \int dx\, \frac{\delta}{\delta f_x}
\left[ P \left. \frac{\delta H}{\delta g_x}\right|_{g=\delta S/\delta f}
\right] . 
\end{equation}
Eqs. (\ref{hj}) and (\ref{hjcont}) generalise immediately to
multicomponent fields, and form the basis of the classical
starting point in the derivation of the quantum equations of motion for
bosonic fields in Sec. III.

\section{Proofs of the Theorem and Corollary in Sec. III}

{\bf Proof of Theorem (Eq. \ref{theorem}):} 
From the causality and exact uncertainty assumptions
one has ${\rm Cov}_x(N) = \alpha(P, \delta P/\delta f_x, f_x)$.  To
avoid issues of regularisation, it is convenient to consider a
position-dependent canonical transformation, $f_x\rightarrow
\Lambda_x^{-1}f_x$, such that $A[\Lambda]:=\exp[\int dx\,\ln
|\det\Lambda_x|]$ is
finite. Then the probability density functional
$P$ and the measure $D\!f$ transform as $P\rightarrow AP$ and
$D\!f\rightarrow A^{-1}D\!f$ respectively, and so the
invariance assumption requires that
\[
\alpha(AP, A\Lambda_x^Tu, \Lambda_x^{-1}w) \equiv
\Lambda_x^T\alpha(P,u,w)\Lambda_x , \]
where $u^a$ and $w^a$ denote the vectors $\delta P/\delta f^a_x$ and
$f^a_x$ respectively, for a given value of $x$.
Since $\Lambda_x$ can remain the same at a given point $x$ while varying
elsewhere, this homogeneity condition must hold for $A$ and $\Lambda_x$
independently.  Thus, choosing $\Lambda_x$ to be the identity matrix at
some point $x$, one has $\alpha(AP, Au,w) = \alpha(P,u,w)$ for all $A$,
implying that $\alpha$ can involve $P$ only
via the combination $v:=u/P$.
The homogeneity condition for $\alpha$ therefore reduces to
\[
\alpha(\Lambda^Tv, \Lambda^{-1}w) = \Lambda^T\alpha(v,w)\Lambda~.\]
Note that this equation is linear, and invariant under multiplication of
$\alpha$ by any function of the scalar $J:=v^Tw$.  Moreover, it may
easily be checked that if $\sigma$ and $\tau$ are solutions, then so are
$\sigma \tau^{-1}\sigma$ and $\tau\sigma^{-1}\tau$. Choosing the two
independent solutions $\sigma=vv^T$, $\tau=(ww^T)^{-1}$, it follows that
the general solution has the form
\[
\alpha(v,w) = \beta(J)vv^T + \gamma(J)(ww^T)^{-1} \]
for arbitrary functions $\beta$ and $\gamma$.
For $P=P_1P_2$ one finds $v=(v_1,v_2)$, $w=(w_1,w_2)$, where the
subscripts label corresponding subsystem quantities, and hence the
independence assumption reduces to the requirements
\[
\beta(J_1+J_2)v_1v_1^T + \gamma(J_1+J_2)(w_1w_1^T)^{-1} =
\beta_1(J_1)v_1v_1^T + \gamma_1(J_1)(w_1w_1^T)^{-1}, \]
\[
\beta(J_1+J_2)v_2v_2^T + \gamma(J_1+J_2)(w_2w_2^T)^{-1} =
\beta_2(J_2)v_2v_2^T + \gamma_2(J_2)(w_2w_2^T)^{-1} ,\]
for the respective subsystem covariance matrices.
Thus $\beta=\beta_1=\beta_2=C$, $\gamma=\gamma_1=\gamma_2=D$ for
universal (i.e., system-independent)
constants $C$ and $D$, yielding the general form
\[
\left[{\rm Cov}_x(N)\right]^{ab} = C (\delta P/\delta f^a_x) (\delta
P/\delta f^b_x)/P^2 + D W^{ab}_x[f] \]
for the fluctuation covariance matrix, where $W_x[f]$ denotes the
inverse of
the matrix with ${ab}$-coefficient $f^a_xf^b_x$. Note that
the latter term is
purely a functional of $f$, and hence merely contributes a classical
additive potential term to the ensemble Hamiltonian of Eq.
(\ref{hprime}).  It
thus has no nonclassical role, and can be absorbed directly into the
classical potential $\langle V\rangle$ (indeed, for fields with more
than one
component this term is singularly ill-defined, and hence can be
discarded on physical grounds).  Thus we may take $D=0$ without loss of
generality. Finally, the positivity of $C$ follows from the positivity
of the covariance matrix ${\rm Cov}_x(N)$, and the theorem is proved.
 
{\bf Proof of Corollary (Eq. \ref{corollary}):}
First, the equations of motion corresponding
to the ensemble Hamiltonian $\tilde{H}'$ follow via the theorem and
Eqs.~(\ref{mconj}) as: (a) the continuity
equation Eq. (\ref{mcont}) as before (since the additional term does not
depend on $S$), which from Eq. (\ref{hquad}) has the explicit form
\begin{equation} \label{modcont}
\frac{\partial P}{\partial t} + 2\sum_{a,b}\int
dx\,\frac{\delta}{\delta f^a_x}\left( PK^{ab}_x\frac{\delta S}{\delta
f^b_x}\right) = 0 ;
\end{equation}
and (b) the modified Hamilton-Jacobi equation
\[ \partial S/\partial t = -\delta\tilde{H}'/\delta P = -H[f,\delta
S/\delta f,t] - \delta(\tilde{H}'-\tilde{H})/\delta P .\]
Calculating the last term via Eq. (\ref{theorem}) and
Eq. (\ref{mform}) of Appendix A, this
simplifies to
\begin{equation} \label{modhj}
\frac{\partial S}{\partial t}+H[f, \delta S/\delta f,t]
-4CP^{-1/2} \sum_{a,b}
\int dx\,\left(K^{ab}_x\frac{\delta^2 P^{1/2}}{\delta f^a_x \delta
f^b_x}
+\frac{\delta K^{ab}_x}{\delta f^a_x}\frac{\delta P^{1/2}}{\delta f^b_x}
\right) =0 .
\end{equation}
Second, writing $\Psi=P^{1/2}\exp(iS/\hbar)$, multiplying each side of
Eq. (\ref{corollary})
on the left by $\Psi^{-1}$, and expanding, gives a complex equation for
$P$ and $S$. The imaginary part is just the continuity
equation of Eq. (\ref{modcont}), and the real part is the modified
Hamilton-Jacobi equation of Eq. (\ref{modhj}) above, providing that one
identifies $C$ with $\hbar^2/4$.


\begin{thebibliography}{99}
\bibitem{footact} Note that Eqs. (\ref{mconj}) follow from
the
action principle $\delta \tilde{A}=0$, where the action $\tilde{A}$ is
given by $\int dt [-\tilde{H} + \int D\!f\,S (\partial P/\partial t)]$
(i.e., by the time integal of an ensemble Lagrangian).
\bibitem{footnoteb} In the Hamiltonian formulation the associated
continuity equation for $P[f,g]$ is $\partial P/\partial t + \int dx\,\{
(\delta/\delta f_x)[P(\partial f_x/\partial t)] + (\delta/\delta
g_x[P(\partial g_x/\partial t)]\} = 0$, which reduces to the Liouville
equation $\partial P/\partial t = \{H, P\}$ via Eqs. (\ref{conj}).
\end{thebibliography}
\end{document}